\documentstyle[prl,aps,preprint,amsfonts,amssymb]{revtex}

\newcounter{saveeqn}%
\newcommand{\alpheqn}{\setcounter{saveeqn}{\value{equation}}%
\stepcounter{saveeqn}\setcounter{equation}{0}%
\renewcommand{\theequation}
      {\mbox{\arabic{saveeqn}\alph{equation}}}}%
\newcommand{\reseteqn}{\setcounter{equation}{\value{saveeqn}}%
\renewcommand{\theequation}{\arabic{equation}}}%

\tightenlines

\begin{document}

\title{Suppression of Superconductivity in Mesoscopic Superconductors}
 
\author{M.B. Sobnack and F.V. Kusmartsev}

\address{Department of Physics, Loughborough University, Loughborough,
Leicestershire LE11 3TU, United Kingdom}

\date{\today}

\maketitle

\begin{abstract}

We propose a new boundary--driven phase transition associated with
vortex nucleation in mesoscopic superconductors (of size of the order
of, or larger than, the penetration depth). We derive the rescaling
equations and we show that boundary effects associated with vortex
nucleation lowers the conventional transition temperature in
mesoscopic superconductors by an amount which is a function of the
size of the superconductor. This result explains recent experiments in
small superconductors where it was found that the transition
temperature depends on the size of the system and is lower than the
critical Berezinsk\u{i}--Kosterlitz--Thouless temperature.

\end{abstract}


Phase transitions in two dimensions (2D) has been the subject of
long--standing interest. The main reason is that the
Berezinsk\u{i}--Kosterlitz--Thouless (BKT) transition
\cite{Ber-71,Kos-73,Kos-78} must arise in 2D systems like, for
example, superfluids and quasi--2D superconductors. Berezinsk\u{i}
\cite{Ber-71} was the first to show that topological defects such as
vortices may play a significant role in phase transitions. The
fact that the energy required to create a vortex depends on the size
of the system was the main obstacle to this fundamental
discovery. Berezinsk\u{i}, however, recognized that the creation of a
vortex--antivortex (V-A) pair is energetically more favorable, because
the pair energy depends on the separation distance of the pair only. The
mechanism of the BKT transition is: the creation of V-A pairs
decreases the superfluid density, which in turn decreases the
binding energy of the V-A pair. As a result, it becomes easier for
more pairs to be created and again this reduces the superfluid density
further. This renormalization process continues until the screening
effect is large enough for vortices and antivortices to nucleate
freely and spontaneously. Since single vortices destroy the phase
correlation needed for 2D superfluidity, the superfluidity is
destroyed and there is a jump discontinuity in the superfluid density
at the transition.

In three dimensions (3D), the creation of a V-A pair requires an
enormous amount of energy depending on the size of the system and it
was believed that a generalization of the BKT transition to 3D systems
was not possible. However, Williams \cite{Wil} and Shenoy
\cite{She-89} independently recognized that pairs of vortex loops may
play the same role as V-A pairs in 3D. With these ideas, Williams
\cite{Wil} constructed a model of the superfluid transition in
$^4$He. An analogous theory was developed by Shenoy \cite{She-89} to
describe the phase transition of the 3D XY model. More recently,
Kusmartsev \cite{Kus-96} proposed a mechanism for vortex nucleation in
a flow of rotating superfluid $^4$He, based on a mechanism similar to
the BKT transition, where, however, half--vortex rings (HVR's) play a
role similar to V-A pairs. The HVR's penetrate the barrier with the
help of critical fluctuations via the creation of an HVR
``plasma''. This leads to a topological phase transition where the
barrier for vortex nucleation disappears and spontaneous threshold
vortex generation starts. Kusmartsev derived the scaling laws in the
critical region and introduced and estimated the scaling relation and
scaling exponent for the critical velocity.

The properties of a superconductor are expected to change radically
when its size becomes comparable to that of the Cooper pairs since the
creation energy of a vortex is then of the same order as that of a V-A
pair. Recent progress in microfabrication techniques has made it
possible to study experimentally (mesoscopic and nanoscopic)
superconducting samples of micrometer and nanometer dimensions
\cite{Mos-95,Gei-97,Gei-98}. This has led to the discovery of new
phenomena in the superconductivity of mesoscopic systems, particularly
the discovery of the paramagnetic Meissner effect
\cite{Gei-98,Kus-92,Kus-97,Mos-97}. Moreover, while the value of the
Ginzburg--Landau parameter $\lambda/\xi$ (ratio of the coherence
length $\xi$ to the penetration length $\lambda$) is sufficient to
determine the type of bulk superconductors, both experiments
\cite{Gei-97,Gei-98} and theoretical work
\cite{Deo-97,Sch-98a,Sch-98b,Sch-99a,Sch-99b} on the magnetization of
mesoscopic discs have shown that both the type and order of the
transitions between the superconducting and the normal states depend
on the size of the disc.

In this Letter we propose a boundary--driven phase transition
associated with vortex nucleation in mesoscopic superconductors
\emph{not} previously reported in the literature. The nucleation of
free vortices from the boundary drives the system to a topological
phase transition with a lower critical temperature than the
conventional or BKT critical temperature of the system. Nucleation of
single vortices is prevented by their attraction to the boundary ---
vortices induced in the system, just like 2D electric charges, are
attracted to their mirror antivortices (vortices of opposite
polarity). A single ``free'' vortex can penetrate into the system only
by overcoming the vortex--image antivortex (V-IA) interaction or, in
other words, a surface or Bean--Livingston barrier. However, creation
of other V-IA pairs close to the boundary may renormalize the
``Coulomb'' attraction to the boundary: the V-IA plasma screens the
attraction of the vortex to the boundary just like in the BKT
transition where the creation of V-A pairs screens an effective
interaction between a vortex and an antivortex. Eventually this leads
to the creation of ``free'' vortices. These penetrate into the system
and the order parameter associated with superconductivity is
destroyed.

To illustrate this effect, let us consider a circular flat
superconducting disc of radius $R=O(\lambda)$ and thickness $d\ll R$
and a vortex of vorticity $\kappa=h/m^{\ast}$ ($m^{\ast}$ is the mass
of the Cooper pairs) at a distance $r<R$ from the center of the
disc. Then it is straightforward to show that the image vortex (of
vorticity $-\kappa$) is a distance $r'=R^2/r$ from the center of the
disc on a straight line joining the center to the vortex at $r$ (see,
for example, Ref.~\onlinecite{Gra}). The interaction energy $U_0$ of
the V-IA pair depends logarithmically on the separation $r'-r$,
\begin{displaymath}
U_0=2q^2\ln\frac{R^{\, 2}-r^2}{rr_c} + E_c,
\end{displaymath}
where $E_c$ is the potential energy associated with the core of the
vortex and $r_c$ is the effective core radius. By analogy with the 2D
Coulomb gas,
\begin{displaymath}
q =\left(\frac{\rho_s}{4\pi}\right)^{1/2}\kappa
=\left(\pi\rho_s\right)^{1/2}\left(\frac{\hbar}{m^{\ast}}\right)
\end{displaymath}
is the effective vortex charge. $\rho_s\equiv \rho_s^{\rm
2D}=\rho_s^{\rm 3D}d$ is the 2D superfluid density. $U_0(r)$ is the
energy with which the vortex charge at $r$ is attracted to the
boundary (surface) of the superconducting material.  

At low temperatures, near $T=0\,\text{K}$, it is unlikely that more
than only a few vortices will be present. However at higher
temperatures, there are likely to be many more vortex excitations,
including some located in the space between $r$ and $R$. These have an
attenuating effect on, and screen, the interaction $U_0(r)$. Following
Kosterlitz and Thouless,\cite{Kos-73,Kos-78} and Williams\cite{Wil}
and Shenoy\cite{She-89}, we take into account this screening effect by
introducing a scale-dependent dielectric constant
\begin{equation}
\varepsilon(r)=1+4\pi\chi(r).
\end{equation}
The effective susceptibility
$\chi(r)=\int_{r_c}^{R}\alpha(r')dn(r')$, where
$\alpha(r)=q^2(R-r)^2/2k_BT$ is the polarizability and $n(r)$ is the
number density of vortices. It is straightforward to show that
$dn(r)=2\pi rdr \exp (-U(r)/k_BT)/r_c^4$. $U(r)$ is
the screened interaction,
\begin{displaymath}
U(r)=2q^2\int_{r_c}^{(R^2-r^2)/r}\frac{dr'}
{\varepsilon(r')\,r'} + E_c.
\end{displaymath}
Introducing the dimensionless superfluid density $K=q^2/(\pi k_BT)$
and the renormalized density $K_r=K/\varepsilon(r)$, Eq.~(1) takes
the form
\begin{displaymath}
K_r^{-1}=K^{-1}
+\frac{4\pi^3 y_0}{r_c^4}
\int_{r_c}^{R}dr(R-r)^2\,r 
\exp \left[-2\pi K
\int_{r_c}^{(R^2-r^2)/r}\frac{dr'}
{\varepsilon(r')\,r'}\right],
\end{displaymath}
where $y_0=\exp(-E_c/k_BT)$. This derivation implicitly assumes a
rather low density of vortices. This is evident, for example, in the
fact that we have used the unrenormalized charge $q$ instead of
$q_r=q/\varepsilon(r)$ to determine the polarizability. Next we also
neglect the correction term in the V-IA interaction
energy. Although the principal result is not changed, these
approximations are necessary to prevent the equations from being
intractable and lead to
\begin{equation}
K_r^{-1}=K^{-1}+
\frac{4\pi^3 y_0}{r_c^4}
\int_{r_c}^{R}dr(R-r)^2\,r 
\exp \left[-2\pi K\ln
\frac{R^2-r^2}{rr_c}\right],
\end{equation}
To make Eq.~(2) self--consistent, one needs to replace $K$ in the
exponential by the renormalized density $K_r$. However, at low
temperatures, the integral is small, and Eq.~(2) is the first two
terms in the expansion of $K_r^{-1}$.

At temperatures near the phase transition where the superfluid density
tends to zero, the perturbation series is not valid. In this regime,
we use the vortex-core rescaling technique proposed by Jos\'{e} {\it
et al}.\cite{Jos-77} (see also the review by Wallace \cite{Wal}). The
procedure is to split the range of integration $[r_c,R]$ into two
parts, $[r_c,br_c]$ and $[br_c,R]$, with $b-1\cong \ln b\ll 1$, and
only evaluate the non--singular contribution of small $r$. Rescaling
$r\rightarrow rb$ in the second integral to restore the original
cut--off $r_c$, we find a perturbative expansion for $K_r^{-1}$:
\begin{eqnarray}
\frac{1}{b}K_r^{-1}
&=&
\frac{1}{b}\left[K^{-1} + \frac{4\pi^3}{r_c^4}
r_c^2\,\left(R-r_c\right)^2 y_0
\exp \left(-2\pi K\ln
\frac{R^2-r_c^2}{r_c^2}\right) \ln b\right] \nonumber \\
&+&\frac{4\pi^3 y_0}{r_c^4}
\int_{r_c}^{R/b}dr\left(\frac{R}{b}-r\right)^2\,rb^3
\exp \,(-2\pi K\ln b)
\exp \left[-2\pi K\ln
\frac{R^2/b^2-r^2}{rr_c}\right].
\end{eqnarray}
We now require that Eq.~(3) has the same functional form as
Eq.~(2). This is achieved by introducing new variables at the
increased scale,
\alpheqn
\begin{eqnarray}
K'^{-1}&=&\frac{1}{b}\left[K^{-1}+4\pi^3\frac{(R-r_c)^2}
{r_c^2}y\ln b\right], \\
y'&=&b^3\,y\exp (-2\pi K\ln b), 
\end{eqnarray}
\reseteqn 
together with $R\,'=b^{-1}R$ and $K_r\!=bK_r$, where we have
introduced the fugacity $y=y_0\,\exp \{-2\pi K\ln
[(R^2-r^2)/rr_c]\}$. In deriving the above transformation, we have
only retained terms of $O(y)$. It is convenient to build up a large
increase in the core radius $r_c$ by successive repetition of this
transformation. In this way, one arrives at differential
renormalization group equations for the effective couplings $K_l$ and
$y_l$: 
\alpheqn
\begin{eqnarray}
\frac{dK_l}{dl}&=&K_l-4\pi^3\frac{(R_l-r_c)^2}{r_c^2}K_l^2y_l, \\
\frac{dy_l}{dl}&=&(3-2\pi K_l)y_l, 
\end{eqnarray}
\reseteqn
with the definition $dl=\ln b$. The scaled radius $R_l$ of the
disc satisfies $dR_l/dl=-R_l$.

The fixed point of the rescaling equations is defined by
\begin{displaymath}
(3-2\pi K_l)y_l=0~~~~{\rm and}~~~~K_l-4\pi^3\frac{(R-r_c)^2}
{r_c^2}K_l^2y_l=0,
\end{displaymath}
which have the nontrivial solution 
\begin{equation}
K^{\ast}=\frac{3}{2\pi}~~~~{\rm and}~~~~y^{\ast}=\frac{r_c^2}
{6\pi^2(R-r_c)^2}.
\end{equation}
The critical point ($K^{\ast},y^{\ast}$) separates the two phases of
the system: the first is the low temperature phase, characterized by
growing superfluid (renormalized) density $K_l\approx K_0 e^l$ (with
$K_0$ being the initial value of $K_l$ at scale size $r_c$) and
vanishing fugacity $y_l\approx e^{-l/\xi_0}$, $\xi_0=(\pi
K_0)^{-1}$. The second one is the high temperature phase,
characterized by exponentially growing fugacity: for infinite
temperatures, $K_l=0$ gives $y_l=y_0e^{3l}$, i.e., vortices
proliferate.

The phase transition between these two regimes is easy to understand:
At low temperatures, $T=0{\rm +}\,{\rm K}$, there are only a few
vortices in the system (small fugacity). These are attracted to the
boundary and cannot nucleate. The coherence length $\xi\sim
O(r_c)$. As the temperature increases, there is a growing number of
vortex excitations and these screen the attraction to the
boundary. The superfluid density decreases, the scaled radius $R_l$
decreases and the coherence length increases (as $\sim e^l$). As the
temperature increases further, there comes a point at which the
screening is large enough for vortices to nucleate freely. $R_l\sim
\xi$, the scaling stops and a phase transition occurs.

To find the behaviour of the scaling near the critical point, we
rewrite the rescaling equations (5) in terms of scaled deviations from
the fixed point: we expand $K_l$ and $y_l$ around $K^{\ast}$ and
$y^{\ast}$ as $K_l=K^{\ast}(1+k')$ and $y_l=y^{\ast}(1+y')$. The
scaling equations (5) then become, to first order in $k'$ and $y'$,
\begin{equation}
\frac{d}{dl}
\left(
\begin{array}{c}
k \\
y 
\end{array}
\right)
=
\left(
\begin{array}{cc}
-1 & -1 \\
-3 & \phantom{-}0 
\end{array}
\right)
\left(
\begin{array}{c}
k \\
y 
\end{array}
\right)
\end{equation}
where we have dropped the primes. Expanding $k$, $y$ and $R$ in
eigenstates $A_{\pm}(l)=A_{\pm}e^{\lambda_{\pm}l}$ of the fixed--point
stability matrix above, the eigenvalues are
$\lambda_+=(\sqrt{13}-1)/2$ and $\lambda_-=-(\sqrt{13}+1)/2$. These
define the relevant and irrelevant axes in the $K_l-y_l$ plane. We
assume, following existing procedure, that the relevant scaling field
$A_+$ is the temperature axis, $A_+\approx A|\epsilon|$, where
$\epsilon=(1-T/T_c)$ is the deviation of the temperature $T$ from the
transition (superconducting or Berezinsk\u{i}--Kosterlitz--Thouless
temperature) $T_c$, and $A$ is a constant.

Then the rescaling law for the free energy $F$ per unit area implies
\begin{eqnarray}
Z(K_0,y_0,R_0)
&=&e^{-(F_l-F_0)L^2}Z(K_l,y_l,R_l) \nonumber \\
&=&e^{-(F_l-F_0)L^2}Z\left(A|\epsilon|e^{\lambda_+l},
A_-e^{\lambda_-l},R_0e^{-l}\right),
\end{eqnarray}
where $Z$ is the partition function and $R_0$ is the (unscaled) radius
of the superconducting disc. The scaling stops when $R_l=R_0e^{-l}$
reaches a critical radius $R_c=\xi=r_ce^l$. This happens when
\begin{equation}
l=l_c=\frac{1}{2}\ln \frac{R_0}{r_c}. 
\end{equation}
Setting $l=l_c$ in Eq.~(8), the partition function is well defined
only if
\begin{equation}
\xi(=r_ce^{l_c})=r_c|\epsilon|^{-1/\lambda_+}.
\end{equation}
Combining Eqs.~(9) and (10) gives
$|\epsilon|^2=(r_c/R_0)^{\lambda_+}$, whence
\begin{equation}
T=T_c\left[1-\left(\frac{r_c}{R_0}\right)^{\lambda_+/2}\right].
\end{equation}
This result implies that the superconductivity breaks down at a
temperature which is lower than the conventional superconducting or
BKT critical temperature $T_c$. The mechanism of vortex nucleation
into the disc we have described leads to a depression of $T_c$ by an
amount which varies inversely with the radius of the superconducting
disc
\begin{equation}
\Delta T_c\propto T_c\left(\frac{r_c}{R_0}\right)^{\lambda_+/2}.
\end{equation}
This depression in the critical superconducting/BKT temperature has
been observed in the recent experiments of Geim {\it et al.}
\cite{Gei} on the magnetization of mesoscopic superconducting discs of
various radii, typically with $d\sim 0.1\,\mu\text{m}$ and $R_0\sim$
1--10\,$\mu\text{m}$. For these (aluminium) discs, $r_c$ (equal to the
coherence length $\xi_0$ at zero temperature) can be estimated as
$r_c\sim \xi_0\sim 0.18\,\mu\text{m}$.

The phenomenon described above is very general. Obviously it may arise
in any small system, like superfluid droplets, quantum dots in a
superconducting state and so on. It may arise not only in superfluid
or superconducting systems, but also in other condensed states, such
as the magnetic state. In any case, topological defects originating
from the surface/boundary (like vortices in superfluids and
superconductors) have the potential to destroy a condensed state
provided that the system has a small size and, in general, the
critical temperature must decrease with the size of the system.

We have described a mechanism of vortex nucleation in superconducting
systems of size comparable to the characteristic size (coherence
length) of the quasiparticles of the systems, the Cooper pairs. A
vortex created in a small superconductor is strongly attracted to the
boundary of the system and hence cannot nucleate. The attraction is
due to the Coulomb attraction of the ``vortex charge'' to the ``vortex
charge'' of the image antivortex. However, a fluctuating creation of
several of these topological defects in the bulk of the
superconductor, in the space between the vortex and the boundary,
screens the V-IA attraction by renormalizing the superfluid
density. This further improves the condition for the creation of more
of such fluctuations. The screening effect of the plasma of vortex
fluctuations continues until vortices nucleate freely. This leads to a
phase transition at which the order parameter associated with the
superconductivity of the system is destroyed.

We have used a real--space renormalization group method to derive the
scaling laws in the vicinity of this phase transition, and shown that
the phase transition occurs at a temperature which is lower than the
superconducting or BKT temperature $T_c$. The amount by which $T_c$ is
lowered is equal to $T_c(r_c/R_0)^a$ (with $a>0$), showing that the
depression in the transition temperature increases with decreasing
disc radius $R_0$, in agreement with recent experiments \cite{Gei} on
mesoscopic superconductors.

The properties of a mesoscopic superconductor are remarkably different
from those of macroscopic or even microscopic superconductors, being
dependent on the size of the system. In this Letter, we have predicted
a new phase transition that occurs in such confined systems and driven
by the boundary of the systems. More precisely, this phase transition,
although similar to the BKT transition, is driven by the creation of a
plasma between the vortex and the boundary which screens the Coulomb
attraction of the vortex to the boundary. In this respect, the driving
mechanism for the phase transition is analogous to the mechanism
proposed by Kusmartsev \cite{Kus-96} for the nucleation of vortices in
a flow of rotating superfluid $^4$He.

We would like to thank A.K. Geim for discussing his recent
experimental results with us prior to publication. We would also like
to acknowledge financial support from the EPSRC. FVK is an associate
fellow of the L.D. Landau Institute for Theoretical Physics, Moscow,
Russia.

\end{document}